# Kinetic Flux Equations for Ion Exchange in Silicate Glasses

Guglielmo Macrelli[1]

[1]*Isoclima SpA – R&D Department*

*Via A.Volta 14, 35042 Este (PD), Italy*

*guglielmomacrelli@hotmail.com*

*Correspondence: guglielmomacrelli@hotmail.com

**Abstract**

Ion exchange kinetic flux equations have been extensively investigated since the mid-twentieth century and continue to provide a fundamental framework for describing mass transport phenomena in solid materials. Despite the maturity of this field, inconsistencies remain in the literature concerning the definition, dimensional consistency, and physical interpretation of the parameters involved. A rigorous and unified treatment of these equations is therefore essential to ensure the reproducibility and comparability of theoretical and experimental studies. The present study aims to establish a coherent and systematic development of ion exchange kinetic flux equations, with particular emphasis on the consistent definition and dimensional formulation of the relevant physical quantities. Beyond refining the theoretical foundations, this study extends the classical formulation by incorporating the influence of mechanical stress on ion transport and considering cross-term interactions within the framework of linear irreversible thermodynamics. This study investigates ion-exchange kinetics within silicate glasses, operating under the Nernst-Planck binary interdiffusion regime. These developments provide a more comprehensive description of ion exchange kinetics, particularly as applied to silicate glasses, where coupling between chemical and mechanical effects plays a crucial role in determining transport behavior and performance.

1. Introduction

Ion exchange processes in silicate glasses have long been recognized as a key modification route for tailoring mechanical, chemical, and optical properties of glass articles[1]. Since the pioneering studies in the mid-twentieth century[2,3], numerous experimental and theoretical efforts [4,5,6,7,8,9] have sought to characterize the diffusion of alkali ions and to quantify the associated fluxes across different glass compositions and thermal conditions. Classical models often treat ion exchange as a diffusion-controlled process governed by Fickian transport, subsequently extended to include electrochemical and thermodynamic driving forces. However, while these frameworks have achieved significant descriptive success, they frequently rely on simplified or inconsistent assumptions regarding the definition of transport fluxes, the interpretation of mobility and diffusivity parameters, and the mutual influence between chemical and mechanical gradients. In the broader context of materials thermodynamics, a rigorous treatment of ion exchange phenomena requires a consistent formulation of flux–force relationships derived from the principles of non-equilibrium thermodynamics. Such an approach allows for the explicit coupling of chemical potential gradients, stress fields, and other cross effects that can play a significant role in systems where volume changes, residual stresses, or composition-dependent structural relaxations occur. Despite the availability of theoretical tools from irreversible thermodynamics, their systematic application to ion exchange in glass remains limited, and discrepancies persist in the literature with respect to the dimensional coherence and physical meaning of the kinetic coefficients used. The present study addresses these gaps by developing a comprehensive and internally consistent formulation of kinetic flux equations for ion exchange in silicate glasses. The work emphasizes the rigorous definition of all physical quantities and units, ensuring compatibility with both classical diffusion theory and the thermodynamics of irreversible processes. Furthermore, the model is expanded to include stress-coupled contributions to ion transport and cross-term effects between chemical and mechanical driving forces. This framework provides a basis for improved interpretation of experimental data, enhances predictability of ion exchange kinetics under complex conditions, and contributes to the broader understanding of coupled transport phenomena in glassy materials. Any physical property of glass articles modified by ion exchange is necessarily related to the change of alkali concentration resulting from the ion exchange process. This is the reason why kinetic flux equations are of fundamental importance to develop a physical model of the process. In this study we follow terminology and definitions already established in the literature[10,11]. The approach outlined by Poling and Houde-Walter[10] in the definition of the kinetics of ion exchange is quite correct. The ion exchange process involves a dopant ion (A) diffusing through a glass host and exchanging with a constituent ion (B). This study assumes an equimolar exchange between two specific alkali species, where the maximum exchangeable quantity is limited by the initial concentration of the alkali ions in the host which is the total mobile ion content $C_T$. In binary interdiffusion they can be identified two approaches[12,13,14,15,16]: the Darken regime and the Nerst-Planck regime. In glass science, the choice between the Nernst-Planck and Darken regimes for modeling interdiffusion depends on how the glass network responds to the differing mobilities of the diffusing species. The Nernst-Planck regime describes interdiffusion where the rigid glass network prevents convective flow (lattice shift).

- Limiting Factor: The interdiffusion rate is limited by the slowest component.

- Mechanism: To maintain charge neutrality and satisfy the zero-net-flux condition, a "diffusion potential" (internal electric field) develops. This field accelerates the slower ions and slows down the faster ones.

- Application: Typical for low-temperature ion exchange (e.g., strengthening glass) and small-scale/nanoscale diffusion where the glass viscosity is high and the network cannot relax.

The Darken regime describes interdiffusion where the glass network or lattice can relax or move to accommodate the difference in atomic fluxes.

- Limiting Factor: The interdiffusion rate is controlled by the fastest component.

- Mechanism: Because the lattice can shift (the Kirkendall effect), the species with higher mobility "pulls" the slower one along via a convective transport term.

- Application: Typical for high-temperature processes or large distance scales where the glass has a lower viscosity (supercooled liquid state), allowing for plastic deformation or vacancy creation/annihilation to relax stresses.

In this study for binary interdiffusion it is assumed a Nesrt-Planck regime. The above assumptions are summarized in Table 1.

Table 1 – Assumptions of this study related to Ion Exchange

| Element considered | Description |
|---|---|
| **Process Definition** $A_{RES.} + B_{GLASS} \rightleftarrows B_{RES} + A_{GLASS}$ | A dopant ion (A) diffuses from a reservoir into a glass and exchanges with an existing constituent ion (B) according to the Nerst-Planck regime. This is a well-established method, often used to strengthen glass or alter its optical properties. |
| **Equimolar Exchange** | The assumption of an equimolar exchange is standard for maintaining charge neutrality within the material (a 1:1 charge basis, as both are alkali species). |
| **Limiting Factor** | The maximum quantity of ions that can be exchanged is indeed limited by the initial concentration of the mobile ions already present in the host material, referred to as the total mobile ion content. |
| **Geometry/coordinates** | Mono dimensional semi-infinite slab usually indicated by "x" symbol. This is justified when the coordinate of diffusion and the diffusion layer is far smaller than other glass article dimensions. |

Ions concentration in this study is expressed in units (mol/cm$^3$) and is indicated by symbol $C_i$ where i= A,B. Concentrations are functions of time (t) and spatial coordinate (x). Non-dimensional concentration is indicated by symbol $\chi_i$ and is the molar fraction defined as follows:

$$\chi_i(x,t) = \frac{c_i(x,t)}{C_T} , i=A,B, \tag{1}$$

$$C_T = C_A(x,t) + C_B(x,t). \tag{2}$$

Based on the above assumptions and definitions the values of the concentration of ions A and B in the host matrix are function of both time and position where their local balance is always constant and limited to the initial value of the concentration of B ions in the host glass matrix which is a fixed initial value that is indicated $C_0$. Hence, we have:

$$C_0 = C_B(x, t=0) = C_T. \tag{3}$$

## 2. Ion Exchange kinetics in the framework of non-equilibrium thermodynamics

Kinetics phenomena such as ion exchange constitute non-equilibrium processes that may be rigorously analyzed through the framework of non-equilibrium thermodynamic theory[17,18]. Within this context, the kinetic characterization of ion exchange is confined in this study to the domain of the linear thermodynamics of irreversible processes [17,18,19].

This theoretical approach is predicated upon the fundamental postulate that the fluxes ($J_i$) of each diffusing ionic species are governed by a linear superposition of all constituent thermodynamic forces ($F_i$). This relationship is formally expressed through phenomenological equations, which account for both direct transport and the coupling effects between ionic flows.

$$J_i = \sum_i L_{ij} F_j. \tag{4}$$

In equation (4) the proportionality factors matrix $L_{ij}$ are called phenomenological coefficients and are subject to the Onsager reciprocity relationships ($L_{ij} = L_{ji}$)[17,18,19]. In the ion exchange model presented here only ions A and ions B are capable of diffusing hence the equations system (4) can be explicitly written:

$$J_A = L_{AA} F_A + L_{AB} F_B, \tag{5i}$$

$$J_B = L_{BA} F_A + L_{BB} F_B. \tag{5ii}$$

The driving forces can be expressed[10] in terms of chemical potentials ($\mu_A$ and $\mu_B$) of the exchanging ions and of an electric potential $\phi$.

$$F_A = -\frac{\partial}{\partial x}\left(\mu_A + qF\phi\right), \tag{6i}$$

$$F_B = -\frac{\partial}{\partial x}\left(\mu_B + qF\phi\right). \tag{6ii}$$

In equations (6) $q$ is the electric charge of exchanged ions (in our case $q=1$ because we are considering monovalent alkali ions) and $F$ is the Faraday constant. The electrical potential can be externally applied or it can be an internal self-generated potential resulting from different mobilities of the exchanging ions.

Equations (6) can be expressed in terms of the corresponding electric field E considering that:

$$E = -\frac{\partial \phi}{\partial x}. \tag{7}$$

The relationships between driving forces and potentials (equations (6)) can be further expressed[10] in terms of ions concentrations recalling that chemical potentials are expressed in terms of concentration through the chemical activity (a):

$$\mu_i = \mu^0 + RT \ln(a_i), \tag{8i}$$

$$\mu_i = \mu^0 + RT \ln(\gamma_i \cdot \chi_i). \tag{8ii}$$

It shall be pointed out, as correctly outlined by Eliaz and Banks-Sills[20], that activity, as any physical quantity expressed in term of a logarithmic function, shall be non-dimensional. It makes no sense to take a logarithm of a dimensional quantity.

The thermodynamic driving forces can be equivalently expressed in terms of a total electrochemical potential. The advantage of this last approach is to introduce a total chemical potential including all physical driving mechanisms and to express the thermodynamic driving force as simple gradient of the total chemical potential. The integration of electrical potential in the total chemical potential and subsequent thermodynamic force is formally expressed in equations (9):

$$\eta_i = \mu_i + qF\phi, \tag{9i}$$

$$F_i = -\frac{\partial \eta_i}{\partial x}. \tag{9ii}$$

As correctly pointed out by Poling and Houde-Walter[10] following an argument already discussed by De Groot[19], the Onsager reciprocal relations $L_{ij}=L_{ji}$ are meaningless in this discussion because this mutual-diffusion depends from a single kinetic parameter. A similar discussion has been carried out by Tagantsev and Ivanenko[21] where they used the Onsager relationships coming to a similar general diffusion equation reached by De Groot[19] without using Onsager relationships. In next section it will be discussed a first order approximation ion exchange kinetic problem based on the assumption that the cross-terms of equation (4) are negligeable and each ion flux can be expressed:

$$J_i = L_{ii} \cdot F_i, \quad i=A,B. \tag{10}$$

In equation (10) the thermodynamic forces are defined by equations (6) and chemical potentials by equations (8). In the chemical potential definition (equations (8)) and in the thermodynamic force definition (equations (6)) we have considered terms related to concentration and electrical potential. Later on it will be considered the effect of residual strain (stress) in the host matrix to the chemical potential and thermodynamic driving force[18,21]. In this respect it will be introduced a total chemical potential including the stress term[20,23]:

$$\eta_i = \mu_i + qF\phi - \Omega_i\sigma \quad ; \quad \sigma = \frac{1}{3} \cdot Trace(\sigma_{ij}), \tag{11}$$

where $\Omega_i$ is the partial molar volume of the diffusing ion and $\sigma$ is the hydrostatic component of the stress tensor which is related to the hydrostatic pressure ($p_H$): $\sigma = -p_H$.

### 3. Kinetics of ion exchange: approximation without cross terms.

As already pointed out kinetics of ion exchange is widely discussed in the literature. It is a non-equilibrium interdiffusion process where the fluxes of the exchanged ions are related to the gradient of the respective electrochemical potentials and they are subject to suitable constraints dictated by Gibbs-Duhem equation[21] and the conservation of overall fluxes. In a first order approximation that do not takes into account: cross terms in the Flux/Forces relationship and the influence of residual stress the flux equations are:

$$-J_i = \beta_i c_i \left( \frac{\partial \mu_i}{\partial x} - FE \right); i=A,B, \qquad (12)$$

where $\beta_i$ is the mobility of the ions in the matrix, $E$ is the electric field and $F$ the Faraday constant. We can express mobility through the self-diffusion coefficient $D_i$ and the phenomenological coefficient $L_{ii}$ according to the Einstein equation:

$$D_i = \beta_i RT, \quad D_i = \frac{RT}{c_i} L_{ii} = \frac{RT}{C_T \cdot \chi_i}. \qquad (13)$$

Kinetic equations and constraints are:

$$-J_A = \frac{D_A}{RT} c_A \left( \frac{\partial \mu_A}{\partial x} - FE \right), \qquad (14i)$$

$$-J_B = \frac{D_B}{RT} c_B \left( \frac{\partial \mu_B}{\partial x} - FE \right), \qquad (14ii)$$

$$J_A + J_B = 0, \qquad (14iii)$$

$$C_A d\mu_A + C_B d\mu_B = 0. \qquad (14iv)$$

The condition (14iii) represents the electroneutrality condition which is the fundamental condition that drive to the Nerst-Planck regime[12,13,14]. while condition (14iv) is a general thermodynamic condition named Gibbs-Duhem equation[21]. Application of conditions (14iii) and (14iv) to flux equations (14i) and (14ii) allows to calculate the electric field generated by the difference of the ion mobilities in the glass matrix:

$$FE = \frac{D_A - D_B}{c_A D_A + c_B D_B} c_A \frac{\partial \mu_A}{\partial x}. \qquad (15)$$

The insertion of (15) into flux equation (14i) results:

$$-J_A = \frac{c_A}{RT} \left[ \frac{D_A D_B (C_A + C_B)}{cCD_A + C_B D_B} \right] \frac{\partial \mu_A}{\partial x}, \qquad (16)$$

making use of equation (9) which connect chemical potential to activity and introducing the relative concentrations:

$$\chi_A = \frac{C_A}{C_A + C_B}; \chi_B = \frac{C_B}{C_A + C_B} \quad , \tag{17}$$

We came to the flux equation:

$$-J_A = \left[\frac{D_A D_B}{\chi_A D_A + \chi_B D_B}\frac{\partial \ln a_A}{\partial \ln \chi_A}\right]\frac{\partial C_A}{\partial x} = C_T \left[\frac{D_A D_B}{\chi_A D_A + \chi_B D_B}\frac{\partial \ln a_A}{\partial \ln \chi_A}\right]\frac{\partial \chi_A}{\partial x} \quad . \tag{18}$$

Defining an interdiffusion coefficient:

$$\tilde{D}_{AB} = \frac{D_A D_B}{\chi_A D_A + \chi_B D_B}\frac{\partial \ln a_A}{\partial \ln \chi_A} = D_{AB} \cdot n \quad , \tag{19i}$$

$$D_{AB} = \frac{D_A D_B}{\chi_A D_A + \chi_B D_B} \tag{19ii}$$

$$n = \frac{\partial \ln a_A}{\partial \ln \chi_A} \tag{19iii}$$

we can write a familiar Fick-Type expression for the flux equation:

$$-J_A = \tilde{D}_{AB}\frac{\partial c_A}{\partial x} = C_T \cdot \tilde{D}_{AB}\frac{\partial \chi_A}{\partial x} \quad . \tag{20}$$

The ion flux equation (equation (20)) is a Fick-type diffusion equation where the diffusion coefficient is concentration dependent. The activity derivative in equation (18) is known as "thermodynamic factor" and it can be expressed in two different forms depending on the expression of activity:

$$a_A = \gamma_A \chi_A, \tag{21i}$$

$$\tilde{D}_{AB} = \frac{D_A D_B}{\chi_A D_A + \chi_B D_B}\frac{\partial \ln a_A}{\partial \ln \chi_A} = \frac{D_A D_B}{\chi_A D_A + \chi_B D_B}\left(1 + \frac{\partial \ln \gamma_A}{\partial \ln \chi_A}\right). \tag{21ii}$$

Or, following Rothmund and Kornfeld[22]:

$$a_A = (\chi_A)^n \quad , \tag{22i}$$

$$\tilde{D}_{AB} = \frac{D_A D_B}{\chi_A D_A + \chi_B D_B}\frac{\partial \ln a_A}{\partial \ln \chi_A} = \frac{D_A D_B}{\chi_A D_A + \chi_B D_B} n \quad . \tag{22ii}$$

In the literature[8], it is defined the thermodynamic factor *n*:

$$n = \frac{\partial \ln a_A}{\partial \ln \chi_A} = 1 + \frac{\partial \ln \gamma_A}{\partial \ln \chi_A} \quad . \tag{23}$$

When *n*=1 the kinetic behavior is called a regular solution behavior.

### 4. Ion exchange kinetics: residual stress effects.

Let's recall the total chemical potential including both electric potential and stress effects (11):

$$\eta_i = \mu_i + qF\phi - \Omega_i \sigma \tag{24}$$

Where $\Omega_i$ is the partial molar volume of diffusing ion "i" and $\sigma$ is the hydrostatic component (1/3 of the stress tensor trace) of the residual stress tensor. Based on the total chemical potential (24) the following flux equations can be stablished:

$$-J_i = \beta C_i \frac{\partial \eta_i}{\partial x} \quad ; \quad i=A,B \quad , \tag{25i}$$

$$-J_i = \beta C_i \left( \frac{\partial \mu_i}{\partial x} - FE - \Omega_i \frac{\partial \sigma}{\partial x} \right); \quad i=A,B \quad . \tag{25ii}$$

Applying again the electroneutrality condition (Nerst-Planck regime):

$$J_A + J_B = 0 \quad , \tag{26}$$

and the Gibbs-Duhem equation:

$$\frac{\partial \mu_B}{\partial x} = -\frac{C_A}{C_B} \frac{\partial \mu_A}{\partial x} = -\frac{\chi_A}{\chi_B} \frac{\partial \mu_A}{\partial x} \quad , \tag{27}$$

after a lengthy but straightforward algebraic calculation we can evaluate the internal electric field due to the different ion mobilities:

$$FE = \frac{\beta_A - \beta_B}{C_A \beta_A + C_B \beta_B} C_A \frac{\partial \mu_A}{\partial x} - \frac{\beta_A C_A \Omega_A + \beta_B C_B \Omega_B}{C_A \beta_A + C_B \beta_B} \frac{\partial \sigma}{\partial x} \quad . \tag{28}$$

The electric field expression (28) can be than inserted in the flux equation (25ii) for ion A. The gradient of the chemical potential is evaluated:

$$\frac{\partial \mu_i}{\partial x} = RT\left[\frac{\partial \gamma_i}{\partial x} + \frac{\partial \ln(\chi_i)}{\partial x}\right] , \qquad (29\text{i})$$

$$\frac{\partial \ln \gamma_i}{\partial x} = \frac{\partial \ln(\gamma_i)}{\partial \ln(\chi_i)}\frac{\partial \ln(\chi_i)}{\partial x} = \frac{1}{\chi_i}\frac{\partial(\chi_i)}{\partial x}\frac{\partial \ln(\gamma_i)}{\partial \ln(\chi_i)}. \qquad (29\text{ii})$$

Finally, inserting (29ii) into (29i) we have:

$$\frac{\partial \mu_i}{\partial x} = \frac{RT}{\chi_i}\frac{\partial \chi_i}{\partial x}\left[1 + \frac{\partial \ln(\gamma_i)}{\partial \ln(\chi_i)}\right] \qquad (30)$$

After a lengthy but straightforward algebraic manipulation, using the definitions (1),(2),(3),(13) and (17), and (30), we arrive to the following kinetic flux equation:

$$-J_A = C_T \frac{D_A D_B}{\chi_A D_A + \chi_B D_B}\left(1 + \frac{\partial \ln \gamma_A}{\partial \ln \chi_\partial}\right)\frac{\partial \chi_A}{\partial x} - \frac{C_T}{RT}\frac{D_A D_B}{\chi_A D_A + \chi_B D_B}\chi_A(1-\chi_A)\Omega_{AB}\frac{\partial \sigma}{\partial x} , \qquad (31)$$

where:

$$\Omega_{AB} = \Omega_A - \Omega_B . \qquad (32)$$

Establishing definitions (33i), (33ii) The kinetic equation (31) can be rewritten as (33iii):

$$D_{AB} = \frac{D_A D_B}{\chi_A D_A + \chi_B D_B} , \qquad (33\text{i})$$

$$n = \frac{\partial \ln a_A}{\partial \ln \chi_A} = 1 + \frac{\partial \ln \gamma_A}{\partial \ln \chi_A} , \qquad (33\text{ii})$$

$$-J_A = C_T D_{AB} n \frac{\partial \chi_A}{\partial x} - \frac{C_T}{RT} D_{AB}\left[\chi_A(1-\chi_A)\right]\Omega_{AB}\frac{\partial \sigma}{\partial x} . \qquad (33\text{iii})$$

Within the limit of classical linear elastic theory[24], the gradient of the hydrostatic stress, in absence of external forces, can be obtained integrating the Beltrami-Michell equation[24]:

$$\frac{\partial \sigma}{\partial x} = -\alpha C_T \frac{\partial \chi_A}{\partial x} , \qquad (34\text{i})$$

$$\alpha = \frac{2E\Omega_{AB}}{9V_m C_T (1-\nu)} , \qquad (34\text{ii})$$

where, E is the Modulus of elasticity of the host glass matrix, ν is its Poisson ratio and *Vm* is the molar volume of the host matrix. Inserting (34i) into (33iii) we can obtain the following kinetic flux equation:

$$-J_A = C_T D_{AB} \left\{ n + K_{AB} \left[ \chi_A \cdot (1 - \chi_A) \right] \right\} \frac{\partial \chi_A}{\partial x} \qquad (35i)$$

$$K_{AB} = \frac{\alpha C_T}{RT} \Omega_{AB} \qquad (35ii)$$

It is remarkable that ion flux equation with stress effects can be reduced to a non-linear (diffusion coefficient concentration dependent) Fick-Type equation (as equation (20)). Ion flux equation with residual stress effects (35i) reduces to kinetics flux equation (20) when $K_{AB}=0$ that is when residual stress effects are negligeable.

## 5. Ion exchange in the framework of linear thermodynamics of irreversible processes: cross terms effect.

Let's go back to section 3 where we established equations (5) which include cross flux terms. We can repeat the derivation of a flux equation like the (20) and (35i) following the same approach by using the above defined equations (5) and conditions (electroneutrality by conservation of fluxes and Gibbs-Duhem equation). After a lengthy but straightforward calculation considering (28) the final flux equation results:

$$-J_A = \left[ \frac{RT}{C_T} \frac{1}{\chi_A \chi_B} \left( \frac{L_{AA} L_{BB} - L_{AB} L_{BA}}{L_{AA} + L_{BB} + L_{AB} + L_{BA}} \right) \right] n \frac{\partial C_A}{\partial x} \qquad (36)$$

Where *n* is the already introduced thermodynamic factor (34), and $C_T = C_A + C_B$. The diffusion coefficients can be conveniently introduced generalizing definition (13):

$$D_{ij} = \frac{RT}{c_T} \frac{L_{ij}}{\chi_j}; \quad i = A,B \ (i \neq j); \quad D_i = \frac{RT}{c_T} \frac{L_{ii}}{\chi_i}; \quad i = A,B \qquad (37)$$

With this definitions equation the flux equation (36) can be written in terms of diffusion coefficients:

$$-J_A = \frac{D_A D_B - D_{AB} D_{BA}}{\chi_A (D_A + D_{BA}) + \chi_B (D_B + D_{AB})} n \frac{\partial C_A}{\partial x} \quad . \quad (38)$$

Incidentally the above flux equation is quite similar to the one derived by Poling and Houde-Walter[10]. The next step is to try a factorization[25] of the interdiffusion coefficient of equation (38). The aim is to make a direct comparison with flux equation (20) derived without taking into account the cross-diffusion terms resulting from the linear irreversible thermodynamics. In order to develop such factorization let's define the interdiffusion coefficient of equation (38) as follows:

$$D^*_{AB} = \frac{D_A D_B - D_{AB} D_{BA}}{\chi_A D_A + \chi_B D_B + \chi_A D_{BA} + \chi_B D_{AB}} n \quad . \quad (39)$$

We can write the following relationship:

$$D^*_{AB} = \tilde{D}_{AB} \cdot \Psi_{Cross}, \quad (40)$$

where we have used the interdiffusion coefficient (21ii) and (22ii) and (33i). The explicit calculation of $\psi_{cross}$ is straightforward and it results:

$$\Psi_{cross}(\chi_A, \chi_B) = \frac{\chi_A D_A + \chi_B D_B}{\chi_A D_A + \chi_B D_B + \chi_A D_{BA} + \chi_B D_{AB}} \left(1 - \frac{D_{AB} D_{BA}}{D_A D_B}\right), \quad (41)$$

based on the above results, equation (38) can be written as a generalization of equation (20):

$$-J_A = \tilde{D}_{AB} \Psi_{cross} \frac{\partial C_A}{\partial x} \quad . \quad (42)$$

After equation (42), we can introduce a new thermodynamic factor $\psi(\gamma_A, \gamma_B)$ which is a function of glass chemical composition according to:

$$\psi(\gamma_A, \gamma_B) = \Psi_{cross}(\gamma_A, \gamma_B) \cdot n \quad (43)$$

As correctly pointed out by Poling and Houde-Walter[10] the cross-terms introduced with the generalization to linear irreversible thermodynamics represents the interaction between the unlike species and, as such, they shall be introduced in the thermodynamic factor as we have suggested with equation (10).

## 6. Ion exchange kinetics: diffusion equation.

The purpose of the kinetics of ion exchange is the determination of the concentration of the incoming ions in the glass host m $C_A(x,t)$. This means being able to write a diffusion equation for the concentration $C_A(x,t)$ that, under suitable boundary conditions, can be solved by analytical or numerical techniques. The starting point to write down a diffusion equation for ion concentrations is to consider a mass conservation equation for the ion flux:

$$\frac{\partial C_A(x,t)}{\partial t} + \frac{\partial J_A(x,t)}{\partial x} = 0, \qquad (44)$$

coupled with a so-called constitutive equation (ion flux kinetic equation) as a Fick-type equation (20) or (35i) or (42) leading to:

$$\frac{\partial c_A(x,t)}{\partial t} - \frac{\partial}{\partial x}\left(D^*_{AB}\frac{\partial c}{\partial x}\right) = 0. \qquad (45)$$

The diffusion equation (45) is a second-order partial derivative non-linear differential equation, its solution (apart from the mathematical conditions for the existence of a solution) requires initial and boundary conditions. In all considered cases:

A) no cross fluxes and no residual stress effects (section 3),

B) no cross fluxes effects with residual stress component (section 4)

C) cross fluxes included and no residual stress effects (section 5).

It has been possible to write a Fick-type kinetic flux equation with a concentration dependent interdiffusion coefficient. These kinetic flux equations are the constitutive equations that, coupled with a mass conservation equation, allows to define the kinetic diffusion equation for ion concentration in the host (glass) matrix. It shall be pointed out that the resulting kinetic concentration equation is non-linear because of the dependence of the interdiffusion coefficient from the ion concentration and, as such, from the spatial coordinate. The kinetic concentration differential equation (45) becomes linear only on the assumption of constant interdiffusion equation.

## 7. The structure of the thermodynamic factor

The thermodynamic factor (n) introduced in equations (19iii) and (33ii)

$$n - 1 = \frac{\partial \ln \gamma_A}{\partial \ln \chi_A}, \tag{46}$$

can be expressed considering sub-regular thermodynamic mixing properties of a binary solution[14]. In general:

$$\frac{\partial \ln \gamma_A}{\partial \ln \chi_A} = \frac{\chi_A(1-\chi_A)}{RT}\left[W_{AB}\left(2\chi_A - 4(1-\chi_A)\right) + W_{BA}\left(2(1-\chi_A) - 4\chi_A\right)\right]. \tag{47}$$

The coefficients Wij are the Margules parameters[14]. In case of "regular solution" $W_{AB}=W_{BA}=W$ and the thermodynamic factor results:

$$\frac{\partial \ln \gamma_A}{\partial \ln \chi_A} = \frac{2W \cdot \chi_A(1-\chi_A)}{RT}. \tag{48}$$

Following this approach, the thermodynamic factor of the interdiffusion coefficient can be conveniently expressed as a function of concentration allowing suitablw approaches to numerical solutions of the diffusion equation for concentration.

## 8. Ion exchange kinetics: conclusion

In the present study we have derived the kinetic flux equations for ion exchange in silicate glasses under different assumptions. In this study a general yet well justified assumption is considered that Ion Exchange as a binary interdiffusion in a Nerst-Planck regime. Interdiffusion of ions in strongly insulating materials with fixed valence oxides, where anions are quite immobile during the diffusion time, prevents any charge separation during interdiffusion This means that we do not have cations vacancy flow which justify the Nerst-Planck regime assumption[15]. Hence, this assumption is a general approach assumed in all considered cases (A,B and C) which are:

A) no cross fluxes and no residual stress effects (section 3),

B) no cross fluxes effects with residual stress component (section 4) and

C) cross fluxes included and no residual stress effects (section 5).

In all examined cases it has been possible to express the ion flux equations as a Fick Type equation with concentration dependent interdiffusion coefficient. Because of that, equations are non-linear and can be numerically solved, after the definition of of initial and boundary conditions. The structure of the interdiffusion coefficient takes into account both residual stress effects and cross effects due to the superposition of the two ion fluxes. The theory has been build-up in a consistent and coherent way based on the assumptions of equimolar 1:1 ion exchange, electroneutrality (that is sum of the two ion fluxes equal to zero) and making use of the Gibbs-Duhem thermodynamic equation.